\documentclass[showpacs,preprintnumbers,amsmath,amssymb]{revtex4}

\usepackage{graphicx}
\usepackage{dcolumn}
\usepackage{bm}
\usepackage{url}
\usepackage{epsfig}
\usepackage{multirow}
\usepackage{tensor}
\usepackage{empheq}
\usepackage{amsmath}
\usepackage{color}
\usepackage{leftidx}
\definecolor{myblue}{rgb}{.8, .8, 1}

\def\be{\begin{equation}}
\def\ee{\end{equation}}
\def\ba{\begin{eqnarray}}
\def\ea{\end{eqnarray}}

\newcommand{\fr}[2]{\frac{#1}{#2}}
\def\eff{\rm{eff}}
\def\eq{\rm{eq}}
\def\fid{\rm{fid}}
\def\m{\rm{m}}
\def\h{\rm{h}}
\def\r{\rm{r}}
\def\b{\rm{b}}
\def\d{\rm{d}}
\def\c{\rm{c}}
\def\w{\textrm{w}}
\def\N{\rm{N}}
\def\D{\rm{D}}

\def\Omo{\Omega_{\rm{m}0}}
\def\Oro{\Omega_{\rm{r}0}}
\def\Ogo{\Omega_{\gamma0}}
\def\Obo{\Omega_{\rm{b}0}}

\def\Oco{\Omega_{\rm{c}0}}

\def\eV{\rm{eV}}

\def\km{\rm{km}}
\def\cm{\rm{cm}}
\def\ss{\rm{s}}
\def\K{\rm{K}}

\def\Mpc{\rm{Mpc}}
\def\MeV{\rm{MeV}}
\def\YP{\rm{Y_{\rm{P}}}}

\def\hMpc{\rm{h\,Mpc^{-1}}}

\newcommand{\ocm}{\omega_{\rm{c}}}
\newcommand{\ob}{\omega_{\rm{b}}}
\newcommand{\og}{\omega_{\gamma}}
\newcommand{\on}{\omega_{\nu}}
\newcommand{\orr}{\omega_{\rm{r}}}
\newcommand{\om}{\omega_{\rm{m}}}

\newcommand{\ode}{\omega_{\rm{de}}}

\newcommand{\rs}{r_{\rm{s}}(z_{\ast})}
\newcommand{\rd}{r_{\d}(z_{\ast})}
\newcommand{\ts}{\theta_{\ss}(z_{\ast})}
\newcommand{\td}{\theta_{\rm{d}}(z_{\ast})}
\newcommand{\dAcz}{d_{\rm{A}}^{(\rm{c})}}
\newcommand{\rsz}{r_{\rm{s}}}

\newcommand{\tsz}{\theta_{\ss}}

\newcommand{\PRD}{Phys.\ Rev.\ D}

\newcommand{\JCAP}{J.\ Cosmol.\ Astropart.\ Phys.}
\newcommand{\MNRAS}{Mon.\ Not.\ Roy.\ Astron.\ Soc.}

\newcommand{\AnA}{Astron.\ Astrophys.}
\newcommand{\ApJ}{Astrophys.\ J.}

\newcommand{\Li}{\rm{Li}}
\newcommand{\He}{\rm{He}}
\def\ga{\mathrel{\raise.3ex\hbox{$>$\kern-.75em\lower1ex\hbox{$\sim$}}}}
\def\la{\mathrel{\raise.3ex\hbox{$<$\kern-.75em\lower1ex\hbox{$\sim$}}}}

\begin{document}

\title{Probing Dark Energy with Neutrino Number}


\author{Seokcheon Lee}
\email[]{skylee@kias.re.kr}
\affiliation{School of Physics, Korea Institute for Advanced Study, Heogiro 85, Seoul 130-722, Korea}

\leftline{KIAS-P14057}


\begin{abstract}
From measurements of the cosmic microwave background (CMB), the effective number of neutrino is found to be close to the standard model value $N_{\eff} = 3.046$ for the $\Lambda$CDM cosmology. One can obtain the same CMB angular power spectrum as that of $\Lambda$CDM for the different value of $N_{\eff}$ by using the different dark energy model ({\it i.e.} for the different value of $\w$). This degeneracy between $N_{\eff}$ and $\w$ in CMB can be broken from future galaxy survey using the matter power spectrum.

\end{abstract}


\maketitle

\section{Introduction}
\renewcommand{\theequation}{1-\arabic{equation}}
\setcounter{equation}{0}

The existence of relic neutrinos is a generic feature of the hot big bang model. This cosmic neutrino has been indirectly measured from the analysis of the cosmic microwave background (CMB) angular power spectrum, as well as primordial abundances of light elements and other cosmological observables \cite{14041740, 14034852, 14013240, 13095383}. Measurements of the Planck satellite CMB alone have led to a constraint on the effective number of neutrino species, $N_{\eff} = 3.36 \pm 0.34$ which is consistent with the stand cosmological model prediction $N_{\eff} = 3.046$ \cite{13035076}. The effect of neutrino properties on cosmological observables are predicted from theory and might be observationally distinguishable from effects of other cosmological parameters \cite{12126267, 0603494}. If there exists extra relativistic species (like sterile neutrinos, sub-eV axions, and {\it etc.}), then one can vary $N_{\eff}$ to include them into the analysis \cite{13035379}. Both cosmological and particle physics observational evidences for the existence of extra neutrino species are still in debate \cite{14072739, 11012755, 13107075, 10071150, 12125226, 13010824, 12126267, 13017343, 13072904}.

Big bang nucleosynthesis (BBN) has emerged as one of the foundation of the hot big bang theory, compounding the Hubble expansion and the CMB \cite{13076955, 9903309}. Compared to other elements in the early universe ($\D$, $\leftidx{^3}{\He}$, and $\leftidx{^7}{\Li}$), the abundance of helium, $\leftidx{^4}{\He}$ is insensitive to the matter density of the Universe, because all neutrons are tied up in helium. Instead, an increase in the number of neutrino, $N_{\eff}$ causes the faster expansion rate of the Universe, therefore more neutrons will survive until nucleosynthesis which leads to an increase in the Helium abundance, $\YP$. This proportional direction of the degeneracy between $N_{\eff}$ and $\YP$ has the orthogonal direction when one considers the equal ratio of the acoustic scale to the diffusion scale. $\YP$ should be decreased as $N_{\eff}$ increases \cite{11042333}. This fact provides strong constraints on both parameters. One needs to observe $\leftidx{^4}{\He}$ from recombination in extremely low-metallicity regions to be found in extragalactic HII regions. $\YP$ is obtained from the extrapolation to zero metallicity but is affected by systematic uncertainties.  Izotov {\it et al.} use both near-infrared spectroscopic observations and optical range ones of high-intensity HeI emission line in 45 low-metallicity HII regions to get \cite{14086953, 13082100}
\be \YP = 0.2551 \pm 0.0022 \label{YpI} \, . \ee
The primordial abundance of $\leftidx{^4}{\He}$ could be appreciated to the zero-metallicity in terms of an extrapolation by a model of chemical evolution of galaxies. An alternative low value using a Monte Carlo Markov Chain technique is reported by Aver {\it et al.} \cite{13090047}
\be \YP = 0.2465 \pm 0.0097 \label{YpII} \, . \ee

There have been great works on the effects of relativistic species quoted as an effective number of neutrino species, $N_{\eff}$ on CMB and large scale structure (LSS) \cite{13084164, 13045981, 11042333}. However, we focus on the degeneracy between the $N_{\eff}$ and the equation of state (eos) of dark energy, $\w$. This degeneracy can be confused with other degeneracies between $N_{\eff}$ and the Hubble parameter, $\h$.

In the next section, we investigate the degeneracy between $N_{\eff}$ and $\w$ on CMB angular power spectrum. We extend this degeneracy on the matter power spectrum in Section 3. In Section 4, we also investigate the degeneracies between $\w$ (or $N_{\eff}$) and $\h$. We draw our conclusions in Section 5.

\section{CMB and $N_{\eff}$}
\renewcommand{\theequation}{2-\arabic{equation}}
\setcounter{equation}{0}
We briefly review the sensitivity of the CMB angular power spectrum to the cosmological parameters to investigate the degeneracy between the effective number of neutrino, $N_{\eff}$ and the equation of the dark energy, w. Let define the present value of the energy density contrast, $\Omega_{i 0} = \rho_{i 0} / \rho_{\rm{cr} 0}$. In addition to dark energy, $i$ can be either the radiation ($\r$) composed of the photon ($\gamma$) and neutrino ($\nu$) or the matter ($\m$) comprised of the cold dark matter ($\c$) and the baryon ($\b$). We define our fiducial model as a flat $\Lambda$CDM with cosmological parameters values as ($\h,\og,\on,\ob,\ocm,\w,A_{S},n_{s},N_{\eff},\YP,\tau$)=(0.6715, $2.4703 \times 10^{-5}$, $1.7094 \times 10^{-5}$, 0.0221, 0.1203, -1, $2.21\times 10^{-9}$, 0.9616, 3.046, 0.25, 0.0927). First, the ratio of odd to even peaks depends on the balance of the gravity and the pressure in a baryon and photon fluid. Thus, if one wants to obtain the same CMB angular power spectrum for different dark energy models, then one needs to fix the ratio of the present energy density of the baryon, $\Obo \h^2 \equiv \ob$ to that of the photon, $\Ogo \h^2 \equiv \og$. Because the present value of the photon energy density is accurately measured ({\it i.e.} the temperature of the photon), one can fix the energy density of the baryon. Thus, we use the same values of $\ob$ and $\og$ for all models. Second, amplitudes of all peaks depend on the matter and the radiation energies equality epoch, $z_{\eq} = \omega_{\m}/\omega_{\r} -1$. Thus, one needs to fix the $z_{\eq}$ for different values of $N_{\eff}$ and this causes changes in the dark matter energy density.
\be \ocm[N_{\eff}] = \og (1 + 0.22711 N_{\eff}) (1+z_{\eq}) - \ob \label{omegac} \, , \ee
where we use $\om = \ob + \ocm$ and $\orr = \og + \on$.
Also the peak location depends on the characteristic angular size of the fluctuation in the CMB as the acoustic scale. It is determined by the sound horizon at the last scattering, $r_{\ss}(z_{\ast})$ and the comoving angular diameter distance, $d_{A}^{(c)}(z_{\ast})$. The acoustic angular size is defined by
\be \tsz[z_{\ast},N_{\eff}, \w, \h] \equiv \fr{\rsz[z_{\ast},N_{\eff},\w,\h]}{\dAcz[z_{\ast},N_{\eff},\w,\h]} \, , \label{thetas} \ee
where $\h$ is defined from the Hubble parameter $H_{0} = 100 \h \km/ \ss / \Mpc$. Both $\rsz$ and $\dAcz$ are a function of the reduced Hubble parameter, $E(z)$
\be E[z, N_{\eff}, \w, \h] = \fr{H}{H_0} = \fr{1}{\h} \sqrt{\om (1+z)^{3} + \orr (1+z)^{4} + (h^2 - \om - \orr) (1+z)^{3(1+\w)}} \label{Ez} \, . \ee Even though, the expression for $E(z)$ given by Eq.(\ref{Ez}) is true only for the constant $\w$, one can extend the consideration for the time varying ones \cite{}. Because $\ts$ is determined from the observation, one can find the relation between parameters ($N_{\eff}$, $\w$, and $\h$) in order to obtain the same value of $\ts$. In this section, we keep the value of $\h$ fixed and investigate the degeneracy between $N_{\eff}$ and $\w$. For the high $l$ acoustic peaks, CMB anisotropies on scales smaller than the photon diffusion length are damped by the diffusion. One needs to consider the mean diffusion distance at the last scattering surface, $r_{\d}(z_{\ast})$ and the angular scale of the diffusion length, $\theta_{\d}(z_{\ast})$. Increasing $N_{\eff}$ leads to the smaller $r_{\d}$ which would decrease the amount of damping. This effect can be compensated by decreasing the Helium abundance $\YP$. In order to obtain the same CMB angular power spectrum for different cosmological models, one needs to fix the ratio between two angular scales $\ts$ and $\td$. From this fact, one can find the relation
\be \fr{\tsz[z, N_{\eff}, \w, \h]}{\td[z, N_{\eff}, \w, \h, \YP]} = \fr{\rs[z, N_{\eff}, \w, \h]}{\rd[z, N_{\eff}, \w, \h, \YP]} \label{tstd} \, . \ee Thus, one can find $\YP[N_{\eff}]$ from the Eq. (\ref{tstd}) in addition to the $\w[N_{\eff}]$ obtained from Eq. (\ref{thetas}). So far we consider both the locations of acoustic peaks and the ratio of the peak amplitudes. Also, one needs to consider the global amplitude of the peaks. This can be adjusted either by adjusting the amplitude of the primordial density field $A_{s}$ or by matching the integrated Sach-Wolfe (ISW) effect. We investigate the both cases.

\begin{figure}
\centering
\vspace{1.5cm}
\begin{tabular}{cc}
\epsfig{file=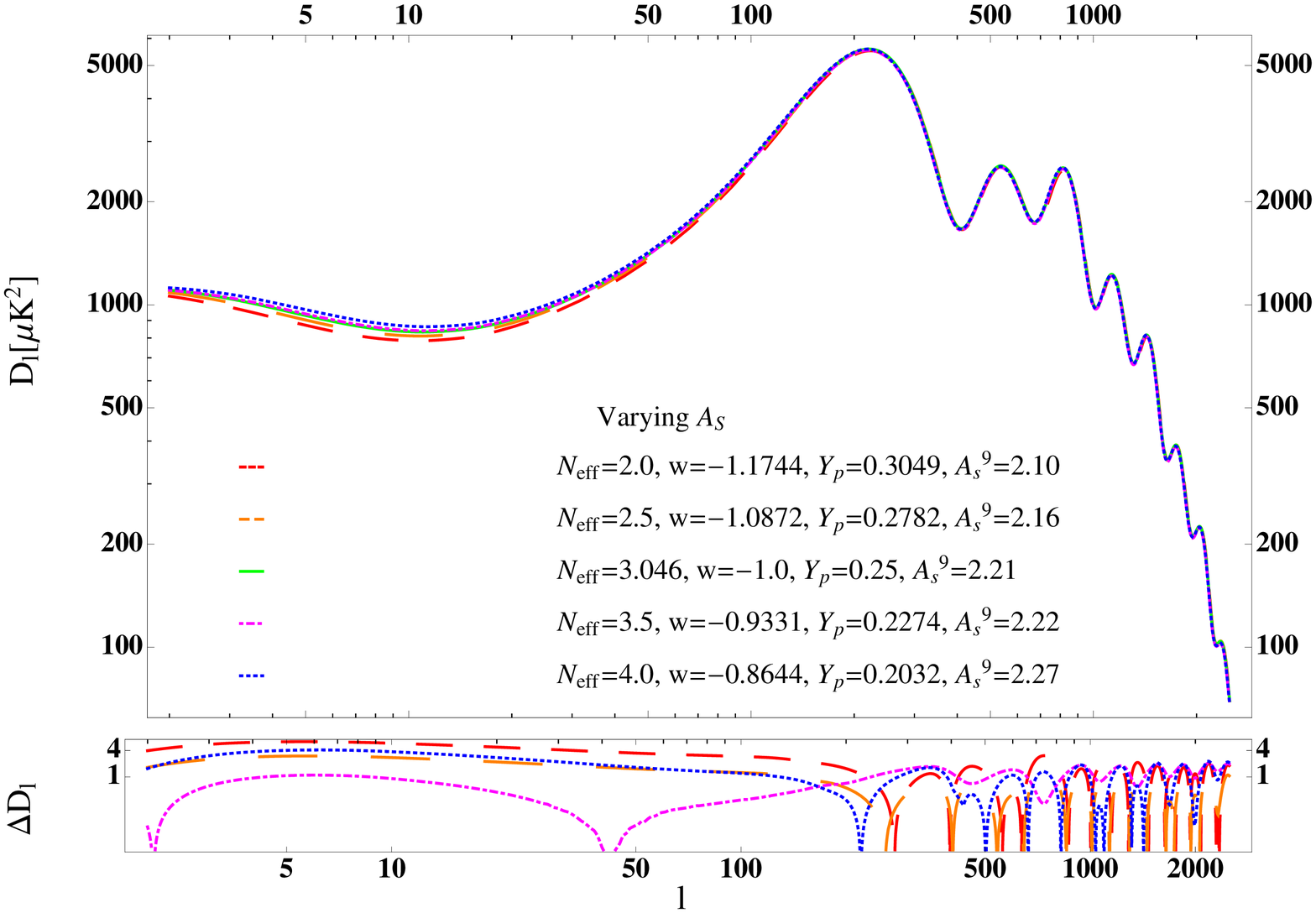,width=0.5\linewidth,clip=} &
\epsfig{file=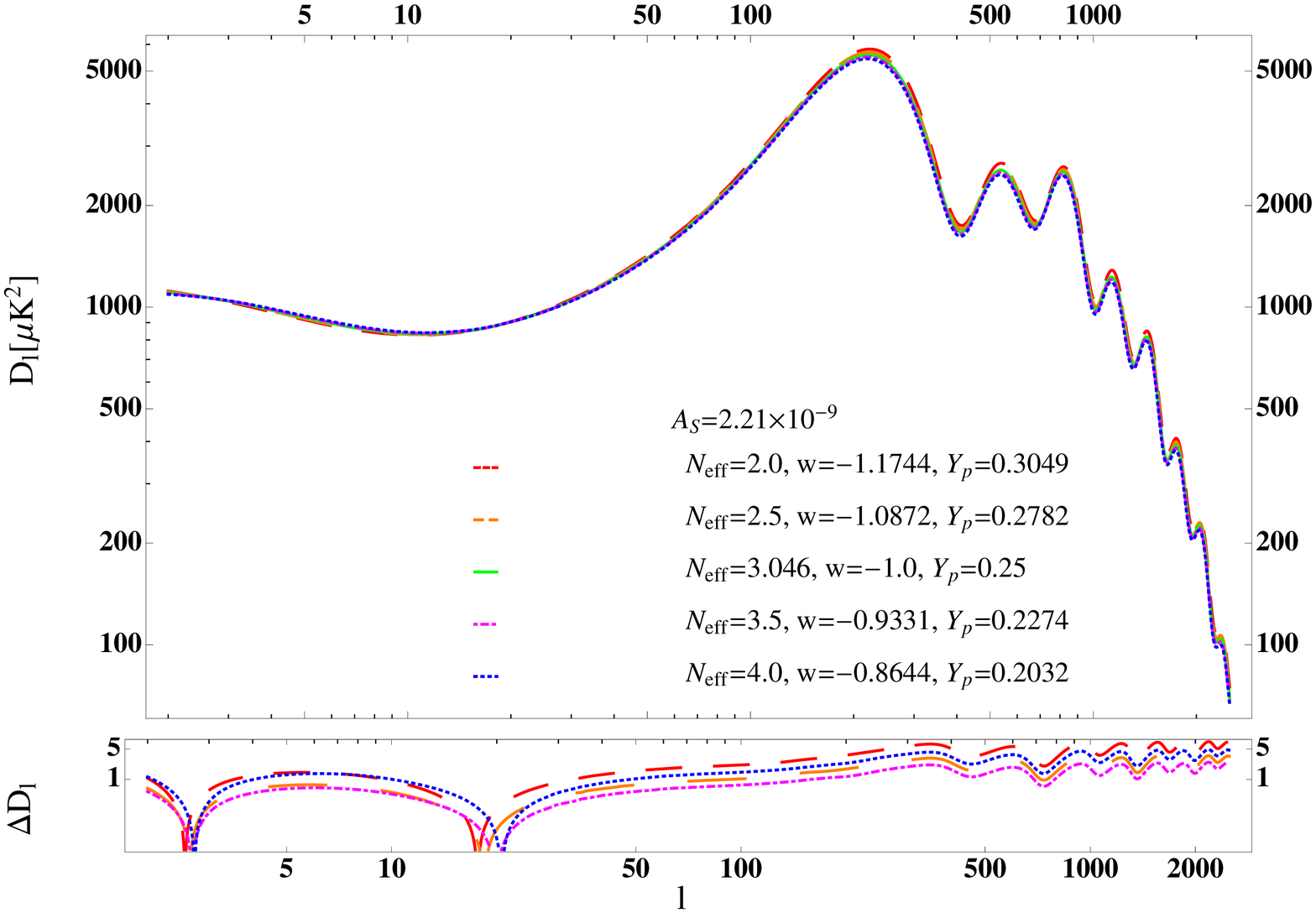,width=0.5\linewidth,clip=} \\
\end{tabular}
\vspace{-0.5cm}
\caption{CMB angular power spectra for different models and their differences from the fiducial model with different normalization. {\it Top left}) CMB angular power spectra for $N_{\eff} = $ 2 (dashed), 2.5 (long-dashed), 3.046 (solid), 3.5 (dot-dashed), and 4 (dotted), respectively. {\it Bottom left}) The differences of CMB power spectra between $N_{\eff} = 2$ (2.5, 3.5, 4.0) model and the fiducial one depicted by dashed (long-dashed, dot-dashed, dotted) line. {\it Top right}) CMB angular power spectra using the same $A_{S}(10^9)$. {\it Bottom right}) The differences of CMB power spectra between models with the same notation as the left panel.} \label{fig1}
\end{figure}

\begin{center}
\begin{table}
\begin{tabular}{ |c||c|c|c|c||c|c|c|c||c|c|c|c|  }
 \hline
$N_{\eff}$ & $\w$ & $\ocm$ & $\ode$  & $\YP$ & $A_{S} (10^9)$ & $\sigma_8$ & $f \sigma_8$ & $\Delta f \sigma_8$ & $A_{S}^{(\rm{fid})} (10^9)$ & $\sigma_8$ & $f \sigma_8$ & $\Delta f \sigma_8$ \\
 \hline
2.0   & -1.1744  & 0.1003 & 0.3285  & 0.3049& 2.11 & 0.837&0.406& -9.1 & 2.21 & 0.856& 0.415 & -7.0\\
2.5   & -1.0872  & 0.1098 & 0.3190  & 0.2782& 2.16 & 0.842&0.426& -4.6 & 2.21 & 0.852& 0.431 &-3.5\\
3.046 & -1.0    & 0.1203 & 0.3086  & 0.25  & 2.21 & 0.845 &0.446& 0& 2.21     & 0.845& 0.446 &0\\
3.5   & -0.9331  & 0.1289 & 0.2999  & 0.2274& 2.22 & 0.840&0.459& 2.8& 2.21   & 0.839& 0.457 &2.5\\
4.0   & -0.8644  & 0.1385 & 0.2903  & 0.2032& 2.27 & 0.841&0.475& 6.5& 2.21   & 0.830& 0.469 &5.1\\
 \hline
\end{tabular}
\caption{CMB degenerated models for the different value of $N_{\eff}$. w, $\ocm (\ode)$, $\YP$ are obtained from Eqs. (\ref{thetas}), (\ref{omegac}), and (\ref{tstd}), respectively. $f$ is the present value of the growth rate of the matter perturbation and $f \sigma_8$ is bias free observable. $\Delta f \sigma_8$ is difference of $f \sigma_8$ between each model and the fiducial one.$A_{S}(10^{9}) \equiv A_{S} \times 10^{9}$ is an input parameter and $\sigma_8$ is obtained from CAMB. $A_{S}^{(\rm{fid})} (10^9)$ is the $A_{S}$ for the fiducial model.}
\label{tab1}
\end{table}
\end{center}
From the above consideration, we obtain various models which can produce almost same CMB angular power spectra as that of our fiducial model. We summarize results in Table.\ref{tab1}. We show the dependence of $\w$, $\ocm$, $\ode$, and $\YP$ on $N_{\eff}$. We also show changes in $\sigma_8$ and $f \sigma_8$ due to the different choice of normalization which will be explained in the next section. As $N_{\eff}$ increases, so does $\on$ when $\ob$, $\og$, $\h$, and $z_{\eq}$ are fixed. This leads to increasing $\ocm$ from Eq. (\ref{omegac}). When $N_{\eff}$ varies from 2.0 to 4.0, $\ocm$ varies from $0.1002$ to $0.1385$. Also as $N_{\eff}$ increases, so does $E(z)$. However, in order to keep $\ts$ equal for increasing $N_{\eff}$, one should increase $\w$ too. This causes the changes in $\w$ from $-1.174$ to $-0.864$ for same ranges of $N_{\eff}$. These values obtained numerically from Eq. (\ref{thetas}). The value of $\YP$ is decreased as $N_{\eff}$ increased in order to keep the same ratio of $\ts / \td$ for different models. One can understand this from Eq.(\ref{tstd}). $\rs$ is decreased as $N_{\eff}$ increases. Thus, one needs to decrease $\rd$ as $N_{\eff}$ and this leads to decreasing $\YP$. $A_{S}$ is inferred in order to reduce the difference of the CMB angular power spectrum at high multipoles between models. Both $\sigma_8$ and $f$ are obtained for given cosmological models. Especially, we adopt $\sigma_8$ values obtained from CAMB. We adopt fiducial model values for the spectral index ($n_{s}$), the optical depth ($\tau$), and the Hubble parameter ($\h$), because $n_{s}$ changes the overall tilt of CMB power spectrum, $\tau$ affects to the relative amplitude for $l \gg 40$ with respect to the lower multipole, and $h$ also affects to the global location and amplitude. We show CMB angular power spectra of different models, $D_{l}$ with different normalizations in Fig. \ref{fig1}. In the top left panel of Fig.\ref{fig1}, we show the CMB power spectra for different models. Their differences between models and the fiducial one with adopting the varying $A_{S}({(9)})$ given in the Table.\ref{tab1} are depicted in the bottom left panel. One can see the degeneracy (less than 2\% for all models) in high $l$ between models. Main differences come from ISW effect which might not be distinguished from the observation. The dashed, long-dashed, solid, dot-dashed, and dotted lines correspond $N_{\eff} =$ (2, 2.5, 3.046, 3.5, 4.0), respectively. In the lower panel, we also show the power spectra difference between various models and the fiducial one, $\Delta D_{l} \equiv \Bigl(D_{l}(N_{\eff})-D_{l}(N_{\eff}^{(\fid)}) \Bigr) / D_{l}(N_{\eff}^{(\fid)}) \times 100 (\%)$ where $N_{\eff}^{(\fid)} = 3.046$. Errors are less than 2 \% for high $l$ except $N_{\eff} = 2$. If one adopt the fiducial model $A_{S} = 2.21 \times 10^{-9}$ for all models, then one obtains almost degenerated CMB power spectra at low $l$. This is depicted in the top right panel. This confirms the fact that ISW effect for the different $N_{\eff}$ values is negligible as shown in \cite{11042333}. The differences of $D_{l}$ are appeared on high $l$. More interesting effect is the shifts on the acoustic peaks. Thus, when one claims the shift in the high $l$ peaks due to the changing in the effective neutrino number, one should also consider the degeneracy in $A_{S}$. The differences become about 5 \% at $l \sim 2000$ as shown in the bottom right panel of Fig.\ref{fig1}.

\section{LSS and $N_{\eff}$}
\renewcommand{\theequation}{3-\arabic{equation}}
\setcounter{equation}{0}
In the previous section, we investigate the CMB angular power spectrum degeneracy between $N_{\eff}$ and other cosmological parameters. It is natural to expect that this degeneracy might be broken in the measurement of the matter power spectrum, $P(k)$. The most obvious effect of varying $N_{\eff}$ appears in the turnover scale at $k \sim 0.02 \hMpc$ which is related to the size of the particle horizon at the matter-radiation equality and hence is determined by $\om$ and $\orr$,
\be k_{\eq}[z_{\eq},N_{\eff},\w,\h] = \fr{H_0}{c} \fr{E[z_{\eq},N_{\eff},\w,\h]}{1+z_{\eq}} \label{keq} \, . \ee
As $N_{\eff}$ increases, so does $E(z_{\eq})$. Thus, $k_{\eq}$ becomes larger as $N_{\eff}$ increases. However, $k_{\eq}$ has yet to be robustly detected and future galaxy survey might provide this information to probe structure at the largest scales. Thus, future galaxy survey is promising to constrain the effective number of neutrino. Also, both the slope and the amplitude of the matter power spectrum depend on the ratio $\ob / \om$. As $N_{\eff}$ increases, so does $\om$ with constant $\ob$. Thus, the slope of the matter power spectrum becomes more moderate as $N_{\eff}$ increases. BAO phase depends on the sound horizon at baryon drag and its amplitude is also related to the Silk damping scale. Thus, one can find the drag epoch of each model if one obtain the accurate BAO signature around $k \sim 0.1 \hMpc$. This effect can provide a useful information on $N_{\eff}$ \cite{14013240}. Also, $\sigma_8$ depends on both $\om$ and $A_s$. If we keep $A_{S}$ fixed, then $\sigma_8$ decreases as $\om$ increases. However, this is not true when we vary the $A_{S}$. The amplitude of the linear matter power spectrum decreases as $N_{\eff}$ increases. Also the growth rate of the matter perturbation, $f$ depends on $\om$. As $\om$ increases, so does $f$. Also one can consider the bias free quantity $f \sigma_8$ which also increases as $N_{\eff}$ does \cite{12056304, 08070810}. The difference in $f \sigma_8$ between models are less than 10 \% as shown in Table.\ref{tab1} and thus can be distinguished from future galaxy survey such as Euclid and LSST.

\begin{figure}
\centering
\vspace{1.5cm}
\begin{tabular}{cc}
\epsfig{file=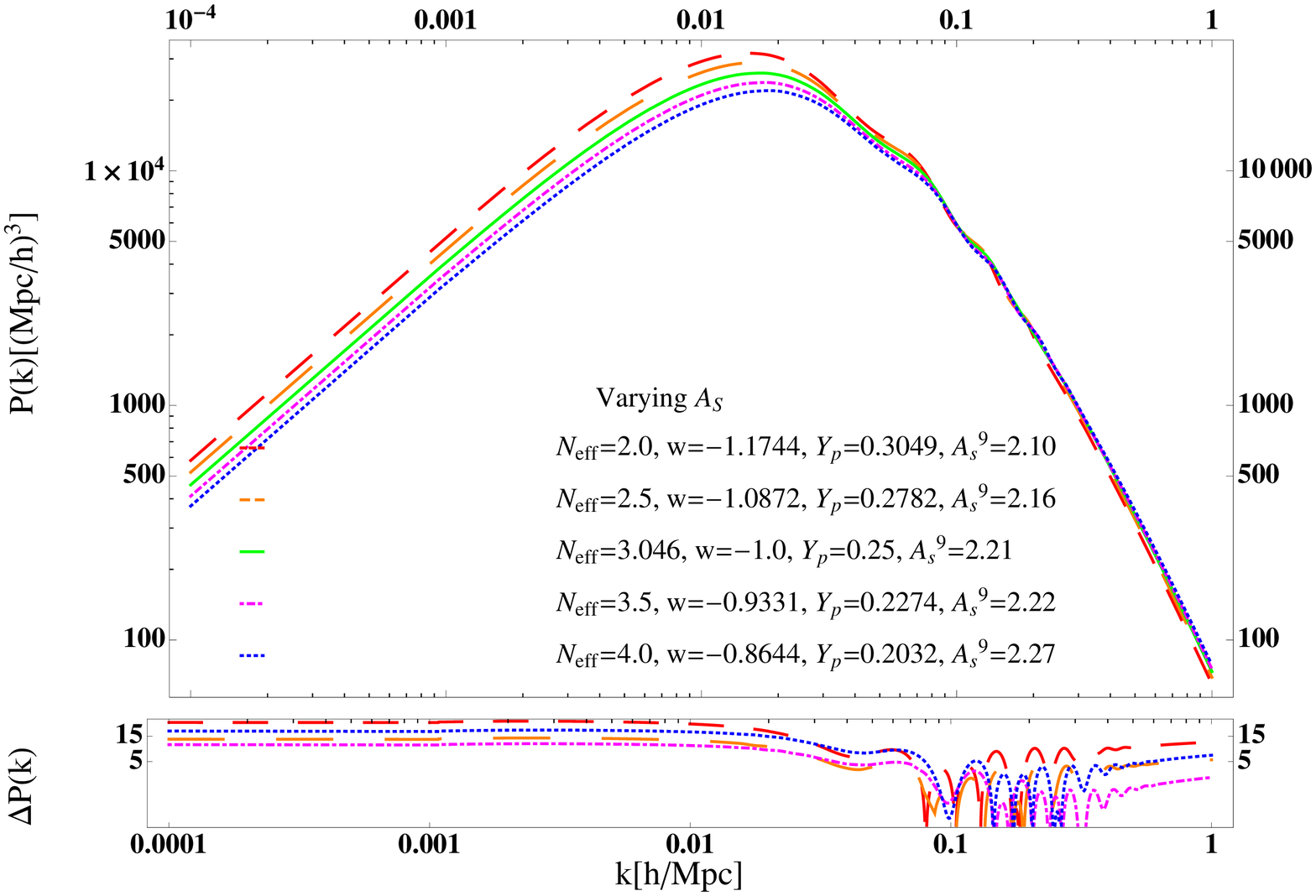,width=0.5\linewidth,clip=} &
\epsfig{file=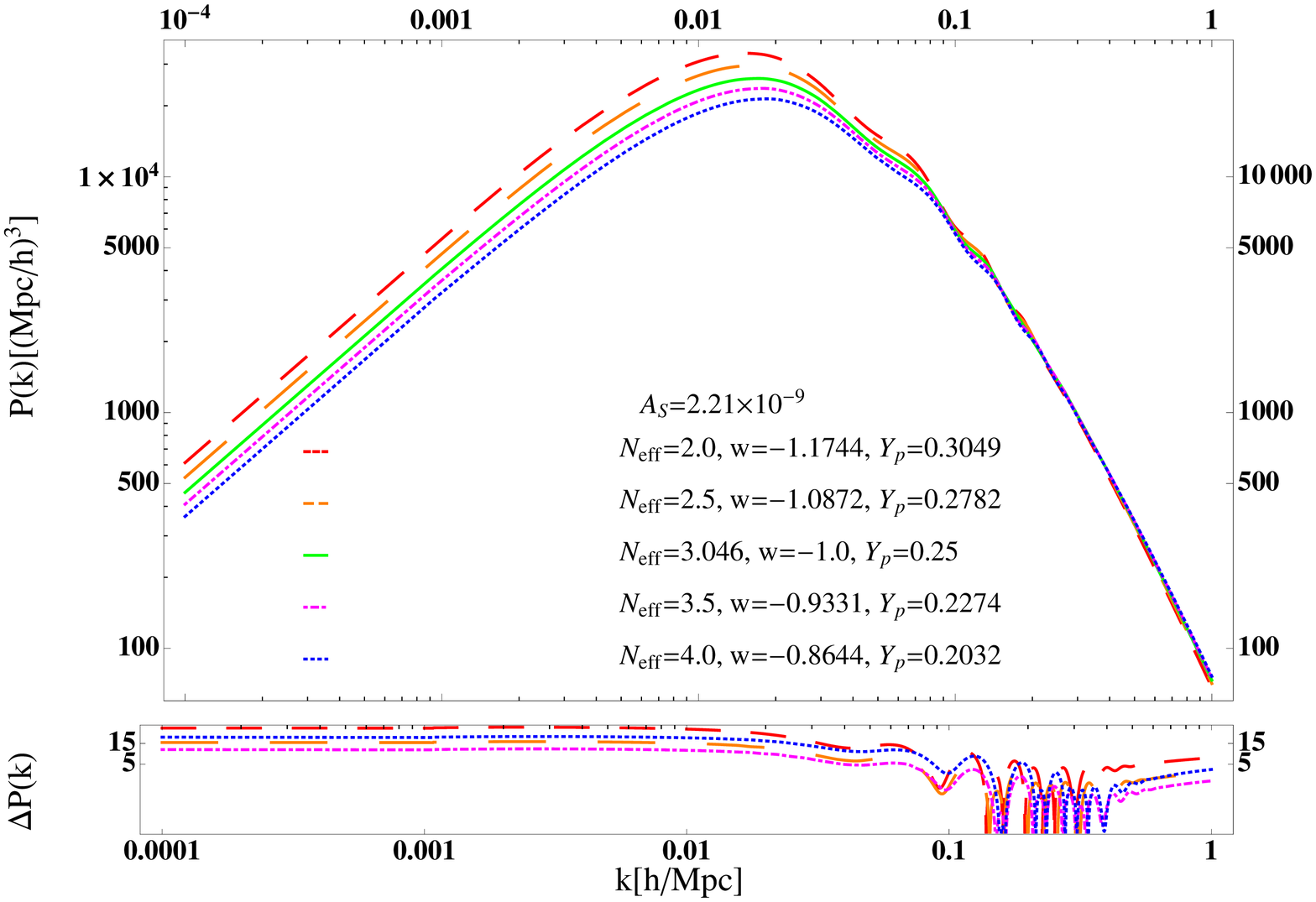,width=0.5\linewidth,clip=} \\
\end{tabular}
\vspace{-0.5cm}
\caption{Linear matter power spectra for different models and their differences from the fiducial model with different normalization. {\it Top left})  CMB angular power spectra for $N_{\eff} = $ 2 (dashed), 2.5 (long-dashed), 3.046 (solid), 3.5 (dot-dashed), and 4 (dotted), respectively. {\it Bottom left}) The differences of CMB power spectra between $N_{\eff} = 2$ (2.5, 3.5, 4.0) model and the fiducial one depicted by dashed (long-dashed, dot-dashed, dotted) line. {\it Top right}) CMB angular power spectra using the same $A_{S}(10^9)$. {\it Bottom right}) The differences of CMB power spectra between models with the same notation as the left panel.} \label{fig2}
\end{figure}

These results are summarized in Table \ref{tab1}. When $A_{S}$ is allowed to vary, there exits no direction of the degeneracy between $\sigma_8$ and $N_{\eff}$ ({\it i.e.} $\ocm$). In this case, $\sigma_8$ becomes $0.837 (0.842, 0.845, 0.840, 0.841)$ when $N_{\eff} = 2.0 (2.5, 3.046, 3.5, 4.0)$. Also $f \sigma_8$ varies from $0.406$ to 0.475 for same ranges of $N_{\eff}$. The direction of the degeneracy between $\sigma_8$ and $\ocm$ becomes the same as the well from the cluster abundance counts, $\sigma_8 \Omega_{m0}^{\gamma} \simeq 0.5$ when $A_{S}$ is fixed. $\sigma_8 (f \sigma_8)$ varies from $0.856 (0.415)$ to $0.830 (0.469)$ for $2.0 \leq N_{\eff} \leq 4.0$. Thus, we should keep in mind the choice of normalization when we claim the CMB and LSS orthogonality of degeneracies between $\sigma_8$ and $\Omega_{\rm{m}0}$. We show the present linear matter power spectra of different models, $P(k)$ with different normalization in Fig.\ref{fig2}. In the top left panel of Fig.\ref{fig2}, we show the matter power spectra for different models. It is obvious that each model has the different turnover scale. $k_{\eq}$ varies from $0.015$ to 0.018 for $2.0 \leq N_{\eff} \leq 4.0$. The slope of $P(k)$ at $k < k_{\eq}$ is same for all models because we fix $n_s$. The dashed, long-dashed, solid, dot-dashed, and dotted lines correspond $N_{\eff} =$ (2, 2.5, 3.046, 3.5, 4.0), respectively. Their differences between models and the fiducial one with adopting the varying $A_{S}({(9)})$ are depicted in the bottom left panel. We define $\Delta P(k) \equiv \Bigl(P(k,N_{\eff})-P(k,N_{\eff}^{(\fid)}) \Bigr) / P(k,N_{\eff}^{(\fid)}) \times 100 (\%)$. $\Delta P(k)$ becomes 25 (18, 12, 10) \% at $k = 0.02 \hMpc$ when $N_{\eff} = 2$ (2.5, 3.5, 4). Also $\Delta P(k)$ is sub-percent level at $k = 0.1 \hMpc$ for all models. If one adopts the fiducial model $A_{S} = 2.21 \times 10^{-9}$ for all models, then one obtains the matter power spectra with more consistent slopes at $k > k_{\eq}$. This is depicted in the top right panel. The differences of $D_{l}$ are appeared on high $l$. In the bottom right panel, we show the $\Delta P(k)$ for the different models. $\Delta P(k)$ becomes 25 (17, 12, 8) \% at $k = 0.02 \hMpc$ when $N_{\eff} = 2$ (2.5, 3.5, 4). $\Delta P(k)$ becomes 3 (3, 1.5, 1.5) \% at $k = 0.1 \hMpc$ for $N_{\eff} = 2$ (2.5, 3.5, 4) model. However, the linear matter power spectrum is not able to be used directly due to bias problem. Thus, it is better to compare the bias free parameter $f \sigma_8$ as we mentioned.

\section{Degeneracies ($\w$,$\h$) and ($N_{\eff}$, $\h$)}
\renewcommand{\theequation}{4-\arabic{equation}}
\setcounter{equation}{0}

In previous sections, we investigate the degeneracy of $N_{\eff}$ and $\w$ from CMB and LSS. We briefly investigate the degeneracy between $\w$ and $\h$ in this section. We also probe the degeneracy between $N_{\eff}$ and $\h$. In the first case, even though there is no change in $N_{\eff}$, $\w$ is degenerated with $N_{\eff}$ and we want to investigate how it can be separated from the degeneracy with $\h$. Results are summarized in Table.\ref{tab2}.

\begin{center}
\begin{table}
\begin{tabular}{ |c|c|c|c|c||c|c|c|c|c|c|c|  }
 \hline
\multicolumn{5}{|c||}{$N_{\eff} = 3.046$} & \multicolumn{7}{|c|}{$\w = -1.0$} \\
 \hline
$\w$ & $\h$ & $\sigma_{8}$ & $f \sigma_8$ & $\Delta f \sigma_8$ & $N_{\eff}$  & $h$ & $Y_{P}$ & $A_{S} (10^9)$ & $\sigma_8$  & $f \sigma_8$ & $\Delta f \sigma_8$ \\
 \hline
-0.8   & 0.6139  & 0.785 & 0.455& 2.1& 2.0  & 0.6226& 0.3049 & 2.12 & 0.7909 & 0.418& -6.4\\
-0.9   & 0.6422  & 0.815 & 0.451& 1.1& 2.5  & 0.6464& 0.2782 & 2.16 & 0.8174 & 0.431&-3.3\\
-1.0   & 0.6715  & 0.840 & 0.446&    0& 3.046  & 0.6715& 0.25 & 2.21 & 0.8452 & 0.446& 0\\
-1.1   & 0.7012  & 0.875 & 0.442&-1.1& 3.5  & 0.6917& 0.2274 & 2.22 & 0.8607 & 0.454& 1.8\\
-1.2   & 0.7320  & 0.905 & 0.437&-2.1& 4.0  & 0.7132& 0.2032 & 2.26 & 0.8819 & 0.466& 4.3\\
 \hline
\end{tabular}
\caption{The degeneracy between $\w$ (or $N_{\eff}$) and $\h$. In the first case, we fix $N_{\eff} = 3.046$ and check the degeneracy between $\w$ and $\h$. In the second case, we fix $\w = -1.0$ and investigate the degeneracy between $N_{\eff}$ and $\h$.}
\label{tab2}
\end{table}
\end{center}

\subsection{($\w$,$\h$)}
\begin{figure}
\centering
\vspace{1.5cm}
\begin{tabular}{cc}
\epsfig{file=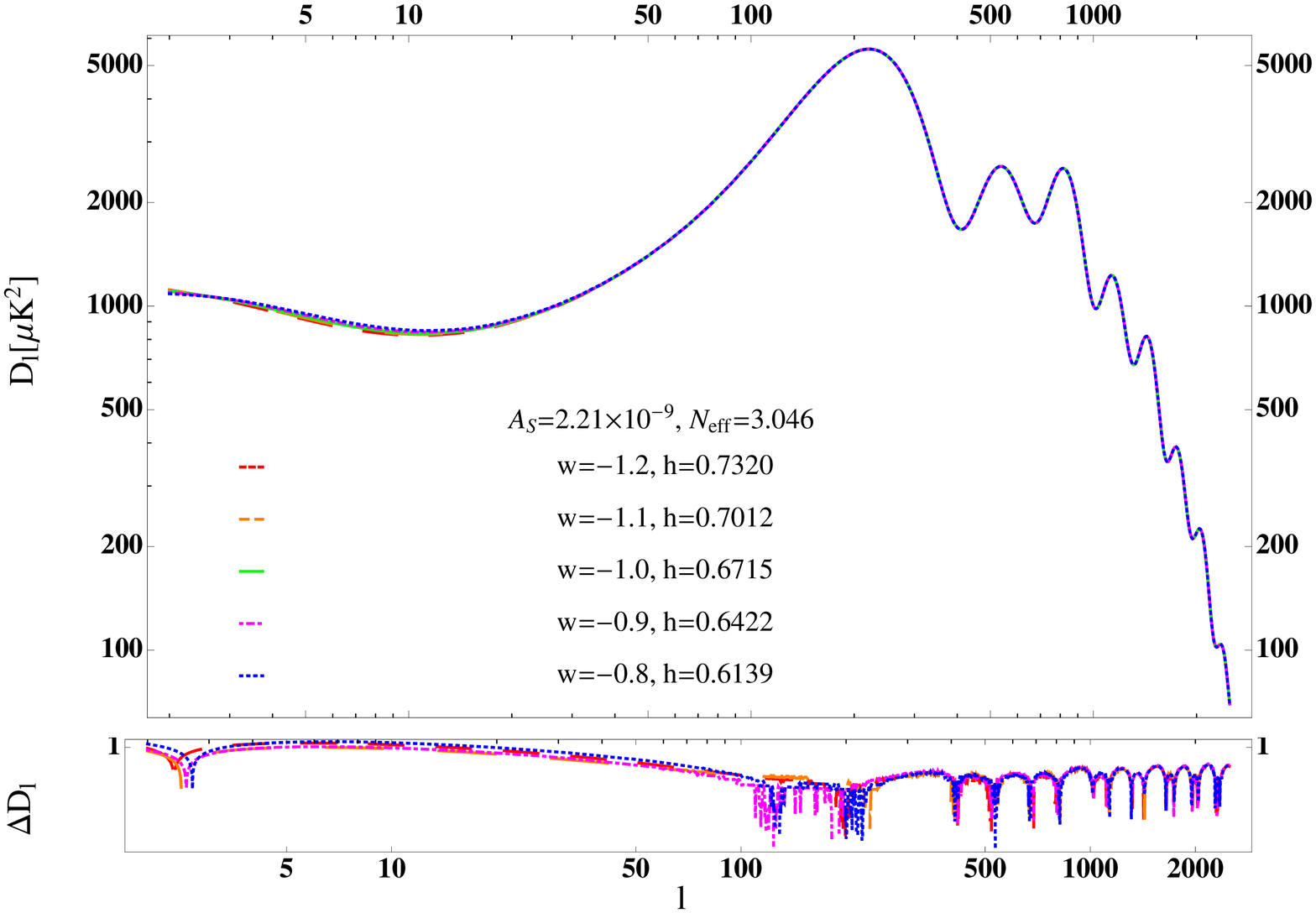,width=0.5\linewidth,clip=} &
\epsfig{file=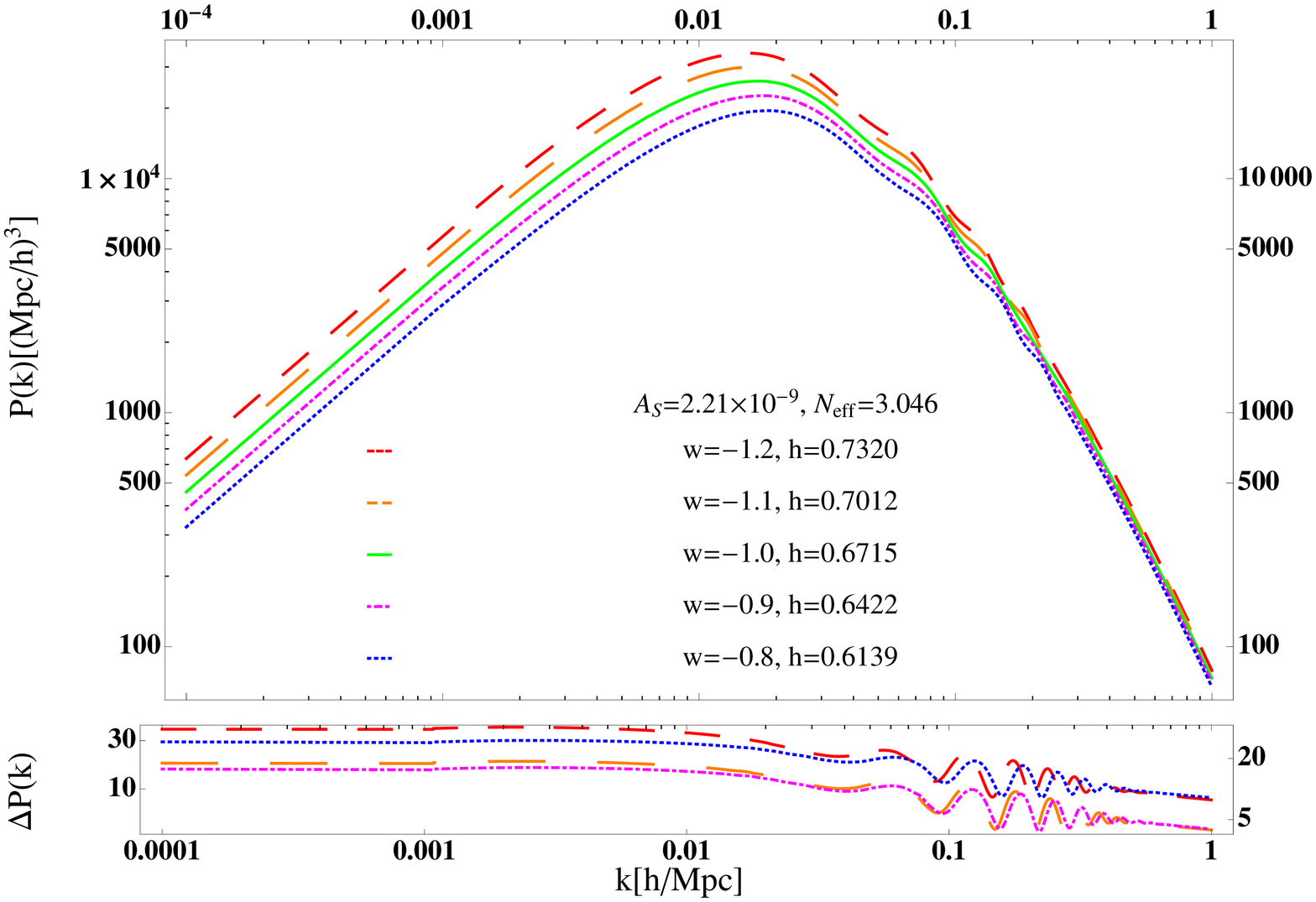,width=0.5\linewidth,clip=} \\
\end{tabular}
\vspace{-0.5cm}
\caption{CMB angular power spectra for different models and their differences from the fiducial model with different normalization. {\it Top left}) CMB angular power spectra for $N_{\eff} = $ 2 (dashed), 2.5 (long-dashed), 3.046 (solid), 3.5 (dot-dashed), and 4 (dotted), respectively. {\it Bottom left}) The differences of CMB power spectra between $N_{\eff} = 2$ (2.5, 3.5, 4.0) model and the fiducial one depicted by dashed (long-dashed, dot-dashed, dotted) line. {\it Top right}) CMB angular power spectra using the same $A_{S}(10^9)$. {\it Bottom right}) The differences of CMB power spectra between models with the same notation as the left panel.} \label{fig3}
\end{figure}

First, we investigate the degeneracy in $\w$ and $\h$ from CMB with $N_{\eff}=3.046$. We keep all other cosmological parameters fixed as a fiducial model. Thus, there is no change in $\ocm$. However, one needs to fix $E(z)$ to produce the same acoustic angular size. If $\h$ increases, so does $\h \times E(z)$ at low $z$. Thus, one needs to decreases $\w$ to make $E(z)$ moderate. Thus, $\h$ decreases as $\w$ increases. If $\w$ varies from $-1.2$ to $-0.8$, then $\h$ changes from $0.7320$ to $0.6139$. This is shown in Table.\ref{tab2}. Thus, increasing $\h$ produces the larger ISW effect at low $l$ in CMB angular power spectrum. Except this effect, models produce almost identical $D_{l}$. This is shown in the top left panel of Fig.\ref{fig3}. The dashed, long-dashed, solid, dot-dashed, and dotted lines correspond $(\w,\h) =$ (-1.2,0.7320), (-1.1,0.7012), (-1.0,0.6715), (-0.9,0.6422), and (-0.8,0.6139), respectively. We use the same normalization $A_{S} = 2.21 \times 10^{-9}$ for all models. The differences between models are about 1 \% at large angle, and they become sub-percent level at high $l$ as shown in the bottom left panel. We also investigate the matter power spectra for models. Because we fix all cosmological parameters except $\w$ and $\h$, the equality wavenumber $k_{\eq} = a_{\eq} H_{\eq} /c$ should be same for all models. However, as one can see in the top right panel of Fig.\ref{fig3}, there are differences in turnover scales between models. This is due to the clustering of the dark energy at large scale ({\it i.e.} at small $k$). On small scales, the dark energy is smooth and the dark energy perturbation is damped and does not contribute the matter density perturbation. However, on large scales, the dark energy clusters and contributes to the energy density and pressure perturbations. The amplitudes of matter power spectra for different models depend on the choice of normalization. We adopt the same primordial amplitude for all models, $A_{S} = 2.21 \times 10^{-9}$. However, one can vary $A_{S}$ and this case the amplitudes of matter power spectra can be changed. When $\w$ decreases, the transfer function of the matter power spectrum. The effective Compton wavenumber of dark energy is approximated as \cite{10102291, 9906174}
\be k_{\w} = \fr{3H}{c} \sqrt{(1-\w)(2+2\w-\w \Omega_{\m}(z))} \label{kde} \, . \ee
Also one can approximate the transfer function as \cite{0309240}
\be T(k) \simeq \begin{cases} 1, & \mbox{when}\,\,\,  k_{\w} D(z_{\eq}) \ll 1 \\ \Bigl(k_{\w} D(z_{\eq}) \Bigr)^{-1}, & \mbox{when}\,\,\, k_{\w} D(z_{\eq}) \gg 1 \end{cases} \label{Tk} \, . \ee
Thus, one find that as $\w$ increases, so does $k_{\w}$. This causes the decreasing $T(k)$ as $\w$ increases. There are about 20 \% differences between models as shown in the bottom right panel of Fig.\ref{fig3}. However, there is bias concern with this differences and if we check the differences of $f \sigma_8$, then the difference between $w= -0.8$ (-0.9, -1.1, -1.2) and the fiducial model becomes 2.1 (1.1, -1.1, -2.1) \% as shown in Table\ref{tab2}. Thus, one needs the percent level accuracy measurement to distinguish the model.

\subsection{($N_{\eff}$, $\h$)}
\begin{figure}
\centering
\vspace{1.5cm}
\begin{tabular}{cc}
\epsfig{file=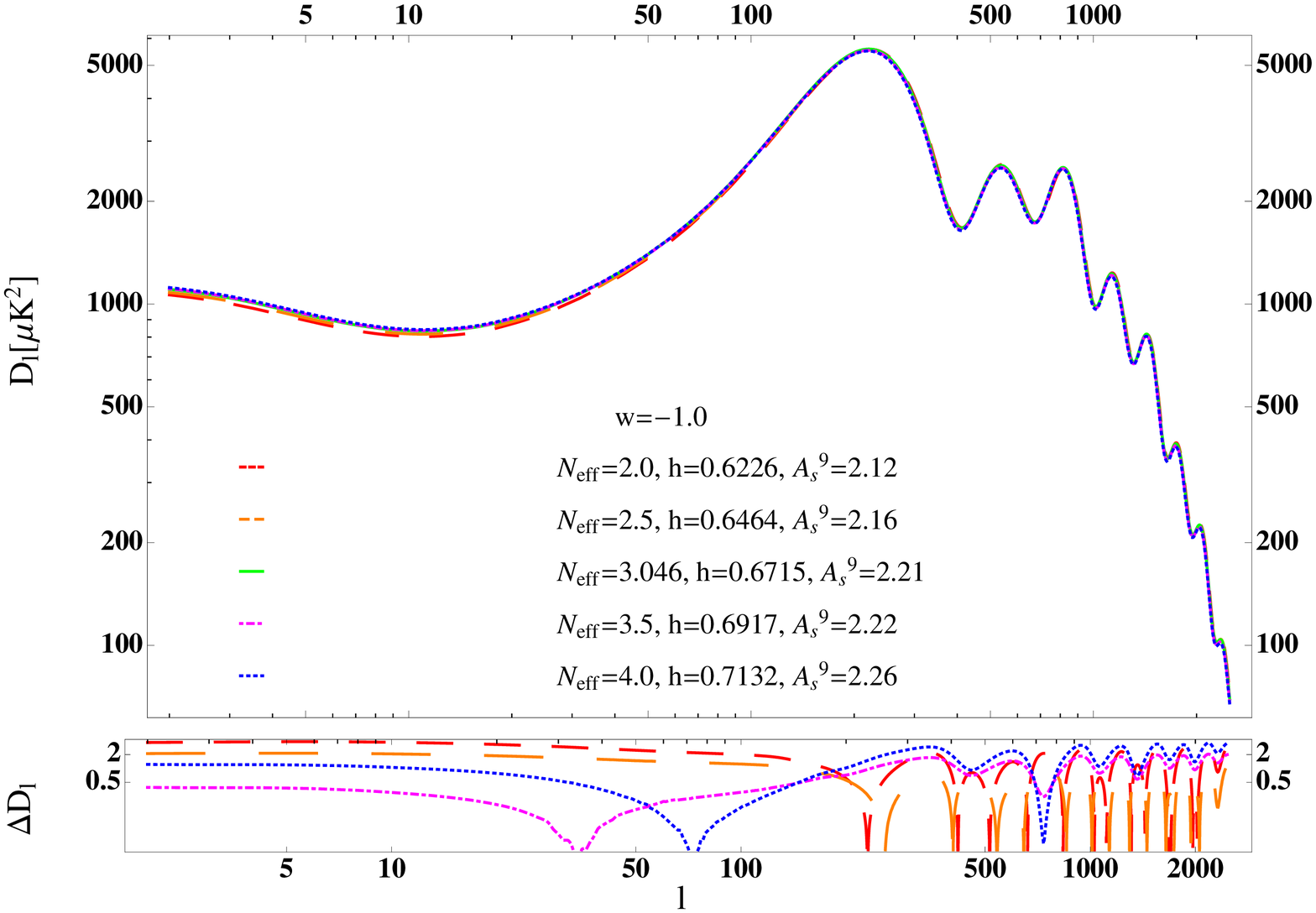,width=0.5\linewidth,clip=} &
\epsfig{file=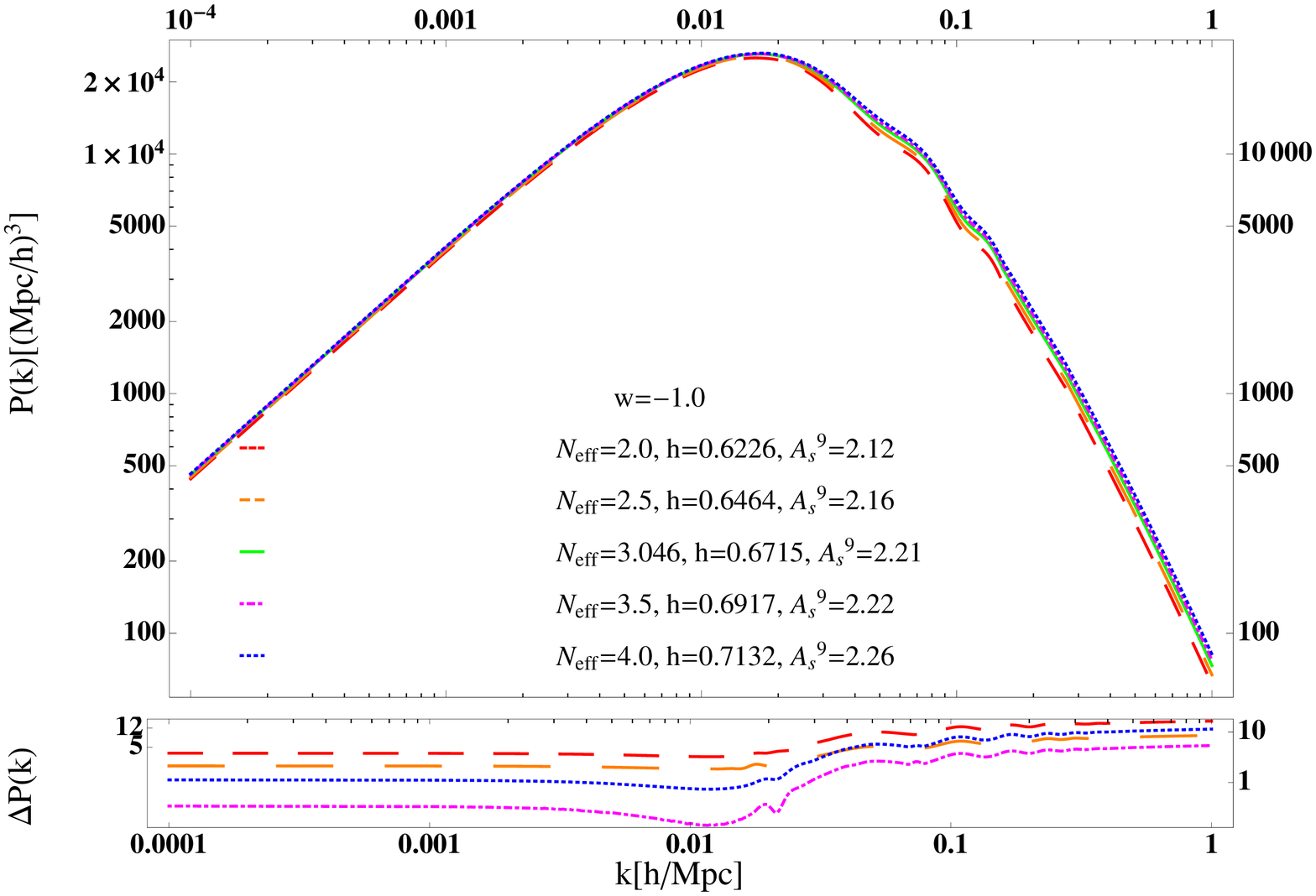,width=0.5\linewidth,clip=} \\
\end{tabular}
\vspace{-0.5cm}
\caption{CMB angular power spectra for different models and their differences from the fiducial model with different normalization. {\it Top left}) CMB angular power spectra for $N_{\eff} = $ 2 (dashed), 2.5 (long-dashed), 3.046 (solid), 3.5 (dot-dashed), and 4 (dotted), respectively. {\it Bottom left}) The differences of CMB power spectra between $N_{\eff} = 2$ (2.5, 3.5, 4.0) model and the fiducial one depicted by dashed (long-dashed, dot-dashed, dotted) line. {\it Top right}) CMB angular power spectra using the same $A_{S}(10^9)$. {\it Bottom right}) The differences of CMB power spectra between models with the same notation as the left panel.} \label{fig4}
\end{figure}

In this subsection, we investigate the degeneracy between $N_{\eff}$ and $\h$. Now, we keep all other cosmological parameters fixed except these two, $\YP$, and $A_{S}$. Again, due to the change in the effective number of neutrino, one obtains the change in the $\ocm$. This also causes the change in the Hubble parameter to match $\theta_{s}$. Also $\YP$ changes due to obtain the same $\theta_{r} / \theta_{d}$ ratio for all models. These changes in $\YP$ and $A_{S}$ are almost same as those in the Section. II, the degeneracy between $N_{\eff}$ and $\w$. In stead of changing $\w$ for the varying $N_{\eff}$, one can obtain almost same effect by varying $\h$. If we fix $\w = -1.0$, $\h$ varies from 0.6226 to 0.7132 when $N_{\eff}$ changes from 2.0 to 4.0. This is shown in Table.\ref{tab2}. The corresponding changes in CMB angular power spectra are dominated in low $l$ due to ISW effect. This is shown in the top left panel of Fig.\ref{fig4}. The dashed, long-dashed, solid, dot-dashed, and dotted lines correspond $N_{\eff} =$ (2, 2.5, 3.046, 3.5, 4.0), respectively. The differences of $D_{l}$ between models are shown in the bottom left panel of the same figure. All are about less than 2 \% for entire region of $l$. The matter power spectra in these models are shown in the top right panel of Fig.\ref{fig4}. As $N_{\eff}$ increases, so does $\h$ and leads to increasing $k_{\eq}$ (inversely decreasing $T(k)$). Thus, one obtains the slight decreasing $P(k)$ as $N_{\eff}$ increases. We use the same notation for this panel as that of the left one. $\Delta f \sigma_8$ is -6.4 (-3.3, 1.8, 4.3) between $N_{\eff} = 2.0$ (2.5, 3.5, 4.0) and the fiducial one. This is shown in the bottom right panel of Fig.\ref{fig4}.

\section{Conclusions}
\setcounter{equation}{0}

We investigate the cosmic microwave background degeneracy on the effective number of neutrino and the equation of state of dark energy. One of the most accurate measurement of CMB is the acoustic scale which depends on both $N_{\eff}$ and $\w$. We showed that CMB is degenerated for the different dark energy models keeping other cosmological parameters fixed \cite{14091355}. This degeneracy might be broken when one combine CMB with LSS. Thus, one should also consider different dark energy models when one investigate the $N_{\eff}$ from CMB and LSS. This will be a challenge for confirming the concordance model. We also investigate the degeneracy between $\w$ (or $N_{\eff}$) and $\h$.

\section*{Acknowledgments}
We would like to thank  for useful discussion. This work were carried out using computing resources of KIAS Center for Advanced Computation. We also thank for the hospitality at APCTP during the program TRP.

\renewcommand{\theequation}{A-\arabic{equation}}
\setcounter{equation}{0}  
\section*{APPENDIX}  

Ratio of odd-to even peaks is due to the gravity-pressure balance in fluid. Thus, we adopt $\fr{\omega_{\b}}{\omega_{\gamma}}$ from Planck. Amplitude of all peaks (damping during MD) depends on $z_{\eq}$. One can find $z_{\eq}$ from the Planck best fit values for $N_{\eff}$ and $\omega_{\c}$. Thus, we fix $z_{\eq}$ for all models. From these, one can directly relate the $N_{\eff}$ to $\omega_c$.  We limit our consideration to the flat universe
\ba \Omo &=& \Oro (1+z_{\eq}) \rightarrow 1+z_{\eq} = \fr{\Omo}{\Oro} = \fr{\Oco + \Obo}{\Ogo \Bigl[1 + \fr{7}{8} (\fr{4}{11})^{4/3}N_{\eff}\Bigr]} \nonumber \\ &\rightarrow& \Oco = (1+z_{\eq}) \Ogo \Bigl(1+0.22711 N_{\eff}\Bigr) - \Obo \label{Oco} \\
&\rightarrow& \omega_{\c}[N_{\eff}] = \omega_{\gamma} \Bigl(1+0.22711 N_{\eff}\Bigr) (1+z_{\eq}) - \omega_{\b} \nonumber \ea
In order to fix the location of peak, one should fix $\theta_{s}(z_{\ast}) = r_{s}(z_{\ast}) / d_{A}^{(c)}(z_{\ast})$. First adopt the best fit value for $N_{\eff} =3.046$, then one uses $\theta_{s\ast} [N_{\eff}, w] = \theta_{s\ast}[3.046,-1.0]$. From this relation, one can find dark energy equation of state $w$ for the different values of $N_{\eff}$. If one fixes $z_{\eq}$, $\og$, and $\ob$, then $\ocm$ depends on $N_{\eff}$ and thus $\w$ depends on $\ocm$ ({\it i.e.} $N_{\eff}$). Now we consider $r_s$ and $r_d$ to make sure the ratio of $\theta_s$ to $\theta_d$ is constant for the different models.
\ba r_s(z_{\ast}) &=& \fr{c}{\sqrt{3} H_0} \int_{z_{\ast}}^{\infty} \fr{dz}{\sqrt{1+R[z]} E[z]} \label{rs} \, , \\
d_{A}^{(c)}(z_{\ast}) &=& \fr{c}{H_0} \int_{0}^{z_{\ast}}\fr{dz}{E[z]} \label{dAc} \, , \\
r_{d}(z_{\ast}) &=& \sqrt{\fr{c \pi^2}{H_0} \int_{z_{\ast}}^{\infty} \fr{(1+z) dz}{\sigma_T X_e n_b (1-Y_P) E[z]} \Biggl[ \fr{R^2 + \fr{16}{15}(1+R)}{6(1+R^2)} \Biggr] } \nonumber \\
&=& \sqrt{\fr{c \pi^2}{H_0} \fr{1}{\sigma_T X_{e} \ob} \fr{1-0.007119Y_P}{(1.12284 \cdot 10^{-5})(1-Y_P)}  \int_{z_{\ast}}^{\infty} \fr{dz}{(1+z)^2 E[z]} \Biggl[ \fr{R^2 + \fr{16}{15}(1+R)}{6(1+R^2)} \Biggr] } \label{rd} \, , \ea
where $R[z] = 3\rho_{b} / 4\rho_{r} = \fr{3}{4} \fr{\ob}{\og} (1+z)^{-1}$, $\sigma_{T}$ is the Thomson scattering cross section, $X_{e}$ is the ionization fraction, and $\YP$ is the Helium fraction.

Big Bang Nucleosynthesis prediction depends on the baryon density $\ob$. It is related to the baryon to photon ratio, $\eta \equiv n_{\rm{b}}/n_{\gamma}$. Relativistic neutrinos contribute to the radiation energy density of the Universe $\rho_{r}$
\be \rho_{\r} = \fr{\pi^2}{15} \Bigl( k_{B} T_{\gamma} \Bigr)^{4}(1+z)^4 \Biggl[ 1 + \fr{7}{8} \, \Bigl(\fr{4}{11}\Bigr)^{4/3} N_{\eff} \Biggr] \label{rhor} \ee
Also the critical energy density of the Universe at present is
\be \rho_{\rm{cr}0} \equiv \fr{3 H_0^2}{8 \pi G_{\N}} = \fr{3 \cdot (100 \h \, \rm{km}/s/ \Mpc)^2}{8 \pi \cdot 6.67191 \times 10^{-8} \cm^3 g^{-1} s^{-2}} \fr{G_{\N}^{(14)}}{G_{\N}} = 1.87901 \times 10^{-29} \h^2 \, \fr{G_{\N}^{(14)}}{G_{\N}} \, [g/\cm^3] \label{rhocr} \, , \ee where we use the new value for the Newton's gravitational constant $G_{\N}^{(14)} = 6.67191 \times 10^{-8} \cm^3 g^{-1} s^{-2}$ \cite{GN}. If one adopts the old value of $G_{\N}$, then one obtains the slightly different value of $\rho_{\rm{cr}0}$. The photon number density is given by
\ba n_{\gamma 0} &=& \fr{2 \zeta(3)}{\pi^2} \Biggl(\fr{k_{B} T_{\gamma 0}}{\hbar c} \Biggr)^{3} = \fr{2 \zeta(3)}{\pi^2} \Biggl( \fr{8.61733 \cdot 10^{-5} (\eV / \K) \times 2.725 (\K)}{4.13567 \cdot 10^{-15} (\eV \cdot \ss)/(2\pi) \times 2.99719 \cdot 10^{10} (\cm / \ss)} \Biggr)^3 \Biggl(\fr{T_{\gamma,0}}{2.725 (\K)} \Biggr)^3 \nonumber \\
&=& 410.802 \Biggl(\fr{T_{\gamma,0}}{2.725 (\K)} \Biggr)^3 \cm^{-3}  \label{ngamma0} \ea
Also, the baryon number density is
\ba n_{B0} &=& \fr{\rho_{B0}}{m_{B}} = \fr{m_{H}}{m_{B}} \fr{\rho_{B0}}{m_{H}} = \fr{m_{H}}{m_{B}} \fr{\rho_{B0}}{\rho_{cr0}} \fr{\rho_{cr0}}{m_{H}} = \fr{1}{1-0.007119Y_{P}} \Omega_{B0} \fr{1.05405 \cdot 10^{-2} h^2 \MeV \cm^{-3}}{938.738 \MeV} \Biggl(\fr{G_{N}^{(14)}}{G_{N}} \Biggr) \nonumber \\
&=& \fr{1.12284 \cdot 10^{-5}}{1-0.007119Y_{P}} \ob \Biggl(\fr{G_{N}^{(14)}}{G_{N}} \Biggr) \cm^{-3} \label{nB0} \, , \ea
where we use $m_{H} = 938.783 \MeV$. Thus,
$\eta_{10} \equiv 10^{10}(n_{B}/n_{\gamma})_{0}$ is given by
\ba \eta_{10} &\equiv& 10^{10} \fr{n_{B0}}{n_{\gamma0}} = \fr{1}{410.802} \fr{1.12228 \times 10^{-5}}{1-0.007119Y_{P}} \Omega_{B0} h^2 \Biggl(\fr{G_{N}^{(14)}}{G_{N}} \Biggr) \Biggl(\fr{T_{\gamma,0}}{2.725 (\K)} \Biggr)^{-3} \nonumber \\
&=& \fr{273.193}{1-0.007119Y_{P}} \ob \Biggl(\fr{G_{N}^{(14)}}{G_{N}} \Biggr) \Biggl(\fr{2.725 (\K)}{T_{\gamma,0}} \Biggr)^{3} \label{eta10} \ea

\end{document}